\documentstyle[prl,aps,preprint,epsfig]{revtex}

\begin{document}

\title{Avalanches, Scaling and Coherent Noise}
\author{M. E. J. Newman}
\address{Cornell Theory Center, Cornell University, Ithaca, NY 14853--3801}
\author{Kim Sneppen}
\address{Nordita, Blegdamsvej 17, DK-2100 Copenhagen \O}
\date{10 June 1996}
\maketitle

\begin{abstract}
We present a simple model of a dynamical system driven by
externally-imposed coherent noise.  Although the system never becomes
critical in the sense of possessing spatial correlations of arbitrarily
long range, it does organize into a stationary state characterized by
avalanches with a power-law size distribution.  We explain the behavior of
the model within a time-averaged approximation, and discuss its potential
connection to the dynamics of earthquakes, the Gutenberg-Richter law, and
to recent experiments on avalanches in rice piles.
\end{abstract}


\section{Introduction}
There has in the last few years been considerable interest in extended
systems which self-organize into a state exhibiting large scale
fluctuations and intermittent dynamics.  One of the earliest attempts to
model systems of this type was made in 1987 by Bak, Tang, and Wiesenfeld,
who proposed a simple lattice model for the avalanches produced by
depositing grains of sand on an ever-growing sand pile~\cite{BTW}.  Despite
having only short-range interactions and no tunable parameters, their model
organizes itself into a state with long-range spatial correlations and
avalanches of size not limited by any finite correlation length.  It has
been proposed that similar self-organized critical (SOC) behavior could lie
behind a wide range of physical phenomena showing $1/f$ noise and
scale-free fluctuation distributions.  SOC models have been put forward to
describe the dynamics of earthquakes~\cite{Quakes}, biological
evolution~\cite{BS} and extinction~\cite{NR}, forest fires~\cite{Fires},
and many other systems~\cite{Review}.  The common features of these models
are that (i)~they are all driven very slowly, and (ii)~they all have
perfect memory, i.e.,~in the absence of the driving force the model would
be entirely stationary.

A distinctive observable consequence of SOC dynamics is that the
distribution of fluctuation (or avalanche) sizes takes a power-law form
with characteristic exponent $\tau$:
\begin{equation}
p_{{\rm aval}}(s) \propto s^{-\tau}.
\label{powerlaw}
\end{equation}
The value of $\tau$ typically lies in the range $1<\tau\le\frac32$, with
the value $\frac32$ corresponding to a critical branching process,
appearing if one makes the ``random neighbor approximation'' in which each
site interacts with a randomly-selected small number of other
sites~\cite{deBoer}.  This approximation is equivalent to the limit of
infinite dimension, and should give correct results for systems above their
critical dimension.  In reality however, a number of the systems which are
modeled using SOC dynamics in fact display event size distributions with
fairly large exponents.  Terrestrial earthquakes, for example, appear to
follow the Gutenberg--Richter law~\cite{GRlaw} with $\tau\approx2.0$, and
the 1D rice pile experiment of Frette and co-workers~\cite{Rice}, which has
been compared to the sandpile model of Bak, Tang and Wiesenfeld, gives
$\tau=2.1\pm0.1$.  A number of models have been proposed which offer
explanations for these higher values of $\tau$.  The model discussed by
Christensen and Olami~\cite{Quakes} is one such, but it achieves its result
only at the price of having an entirely deterministic dynamics; if one
introduces randomness into the model, the simple scaling behavior is
destroyed.  The reason is that when the exponent describing the
distribution of avalanches' spatial extent becomes larger than $2$, the
mean avalanche size becomes finite and independent of system size, and the
spatial overlap between subsequent avalanches becomes
insignificant~\cite{Footnote}.  In the presence of randomness, this can
prevent the system from building up any long-range correlations, and
ultimately destroy the critical state.  We conjecture that, in this regime,
any randomness in the positions of the nucleation centers of the avalanches
will destroy self-organization of long range spatial correlations.

In this paper, we present a different explanation to account for systems
that have larger values of $\tau$.  We demonstrate that power-law event
size distributions having $\tau$ around 2 or greater, are {\em typical\/}
of extended systems with quenched memory if they are driven by coherent
noise, and that in such systems they are present even in the absence of any
interaction between the different parts of the system.  (This is different
from the situation in the SOC models, where the system is driven by a local
driving force, coupled with interactions between the components of the
system.)  The simplest model demonstrating the phenomenon is defined as
follows.  Consider a system of $N$ agents, such as grains on the surface of
a sand pile or points of contact in a subterranean fault.  With each agent
$i$ we associate a threshold for movement $x_i$ which can take values
falling in some specified range and represents the amount of stress that
the agent will withstand before it moves.  For convenience, we choose to
measure $x_i$ on a scale on which $0\le x_i<1$.  The dynamics of the model
then consists of the repetition of two steps:
\begin{enumerate}
\item A fixed fraction $f$ of the agents are selected at random, and the
values of their threshold variables $x_i$ are exchanged with new random
numbers selected uniformly from the interval $[0,1)$.
\item We select a random number or ``stress level'' $\eta$ from some
distribution $p_{{\rm stress}}(\eta)$.  All $x_i$ below $\eta$ are also
exchanged with new random numbers selected uniformly from the interval
$[0,1)$.  The number of agents whose thresholds are changed in this fashion
is the size $s$ of the avalanche taking place in this time-step.
\end{enumerate}

The random selection of different values for $\eta$ at each step may be
thought of as imposing external stresses which {\em coherently\/} (in other
words, simultaneously) influence all of the weaker agents---those having
suitably low thresholds for stress---but leaves unchanged the stronger
ones.  It seems physically reasonable to assume that smaller stresses
should be more common than larger ones, and in the following discussion we
make the assumption that $p_{{\rm stress}}(\eta)$ is largest at $\eta=0$
and falls off to zero as $\eta$ becomes large.  We denote the typical scale
of the falloff by $\sigma$.  The most interesting regime is when $\sigma
\ll 1$ and $f \ll 1$,

\section{Results}
\begin{figure}
\begin{center}
\psfig{figure=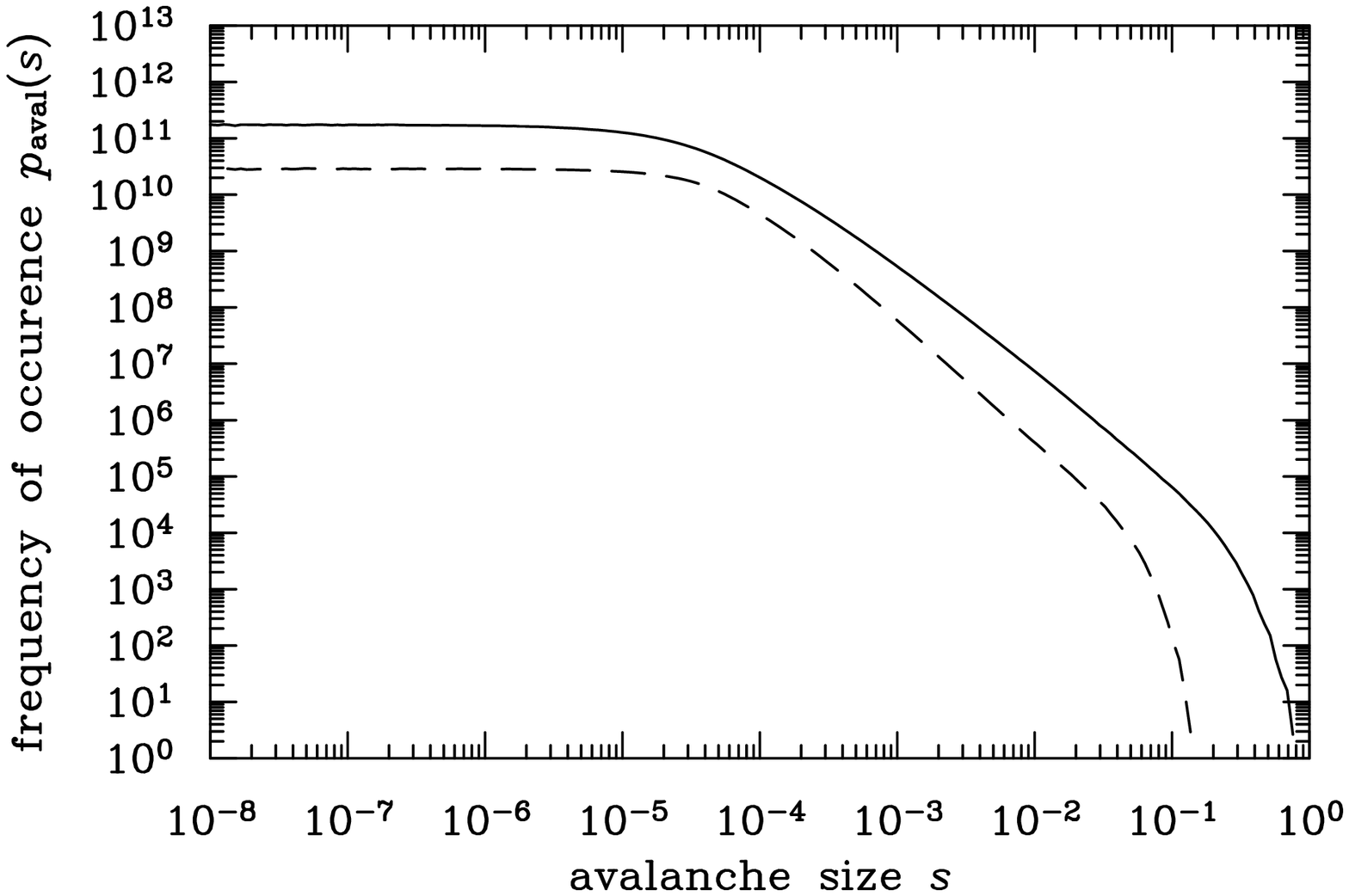,width=13cm}
\end{center}
\caption{Simulation results for the frequency distribution of avalanches
with exponentially distributed stresses (solid line) and Gaussian ones
(dashed line), with $f=10^{-3}$ and $\sigma=\frac1{20}$ in each case.
\label{size}}
\end{figure}

\begin{figure}
\begin{center}
\psfig{figure=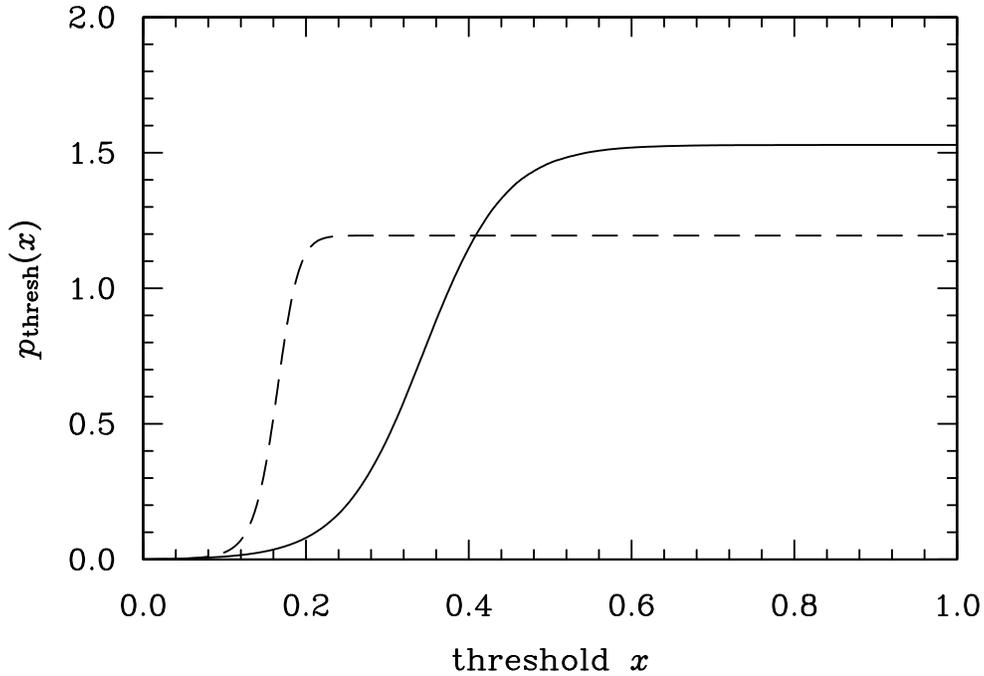,width=13cm}
\end{center}
\caption{Simulation results for the time-averaged distribution of
thresholds $x$ with exponentially distributed stresses (solid line) and
Gaussian ones (dashed line).  As in Figure~\ref{size}, $f=10^{-3}$ and
$\sigma=\frac1{20}$ in each case.
\label{thresh}}
\end{figure}

We have examined the properties of this model both analytically and
numerically. Instead of simulating the model directly, we have developed an
algorithm which calculates the threshold distribution and avalanche sizes
in a formally exact way for a system with $N=\infty$.  Starting off with a
uniform distribution of thresholds, the system evolves towards a
statistically stationary state.  In this state we record the mean threshold
distribution and the frequency distribution of avalanches.  The results are
shown in Figures~\ref{size} and~\ref{thresh} for a simulation using
exponentially distributed stresses:
\begin{equation}
p_{{\rm stress}}(\eta) = \frac1\sigma\exp(-\eta/\sigma).
\label{exponential}
\end{equation}
As Figure~\ref{size} shows, the distribution of avalanche sizes $s$ is flat
up to a certain point (whose position varies with $\sigma$ and $f$) and
then falls off as a power law according to Eq.~(\ref{powerlaw}) with
$\tau\approx2.0$.  This power-law behavior appears to be robust in the
regime of small $f$ and $\sigma$.  If, for example, instead of
Eq.~(\ref{exponential}) we employ a Gaussian stress distribution then,
although the average distribution of thresholds (Figure~\ref{thresh})
changes radically, the power-law form of the avalanche distribution
remains.  Notice, however, that the exponent $\tau$ changes slightly as the
applied stresses are varied.  For the Gaussian distribution, for example,
we find $\tau=2.2\pm0.1$, as opposed to $\tau=1.9\pm0.1$ for the
exponential.  And for steeper distributions of stresses ($p(\eta)\propto
exp(-(\eta/\sigma)^q)$ with $q\ge4$) we find $\tau=2.4$ or greater.

In order to investigate possible connections with spatially-organized
models, we have also implemented our model on a lattice and at each
time-step eliminated not only those agents whose thresholds for stress fall
below the selected level, but also their neighbors.  In all cases we
observe a power-law distribution of avalanches with exponent in the
vicinity of $\tau=2$.

In order to understand the appearance of this power law, let us consider
the time-averaged behavior of the model.  The statistically stationary
state arises as a competition between the two processes comprising the
dynamics: the stresses which tend to remove lower thresholds from the
distribution and thus shift the weight of the distribution to higher values
of $x$, and the aging, which tends to move weight back down again.  The
result is that the average threshold distribution $p_{{\rm thresh}}(x)$ is
a highly nonhomogeneous, monotonic increasing function of $x$ which, for
small $\sigma$, tends to have a plateau as $x$ approaches unity (see
Figure~\ref{thresh}).  By balancing the two competing processes, we can
calculate $p_{{\rm thresh}}(x)$ and hence the avalanche distribution.  For
concreteness, we perform the calculation here for the exponentially
distributed stresses of Eq.~(\ref{exponential}).

The probability of an agent possessing a threshold $x$ lying below the
stress level $\eta$ at any given time-step (and hence of it moving during
this time-step) is
\begin{equation}
P_{{\rm move}}(x) = \int_x^\infty p_{{\rm stress}}(\eta) \>{\rm d}\eta =
{\rm e}^{-x/\sigma}.
\end{equation}
The total time-averaged rate at which agents move in the interval between
$x$ and $x+{\rm d}x$ is then
\begin{equation}
P_{{\rm move}}(x)\, p_{{\rm thresh}}(x) \>{\rm d}x +
f p_{{\rm thresh}}(x) \>{\rm d}x = W \>{\rm d}x
\label{master}
\end{equation}
where the $x$-independent constant $W$ on the right-hand side is the
time-averaged rate at which probability is added to $p_{{\rm thresh}}$.
Rearranging we have
\begin{equation}
p_{{\rm thresh}}(x) = {W\over f + {\rm e}^{-x/\sigma}}.
\label{pthresh}
\end{equation}
The constant is easily fixed by requiring that $p_{{\rm thresh}}(x)$
integrate to unity, giving
\begin{equation}
W = {f\over\sigma} \biggl[ \log{f{\rm e}^{1/\sigma}+1\over f+1} \biggr]^{-1}.
\label{normalization}
\end{equation}
For small $f$ and $\sigma$, $p_{{\rm thresh}}(x)$ rises exponentially
from zero and then levels off in a plateau around $x = -\sigma\log f$.
Physically, this arises because agents possessing thresholds above this
point are affected only by the aging process, which treats them all
equally.  Below this level, the stress process is important too, and it
preferentially moves those with lower thresholds.

The avalanche size distribution is given by
\begin{equation}
p_{{\rm aval}}(s) = \int_0^\infty p(s|\eta)\, p_{{\rm stress}}(\eta) \>
{\rm d}\eta.
\end{equation}
The probability $p(s|\eta)$ of getting an avalanche of a certain size given
a certain stress level, depends on the distribution of thresholds, which
will in general vary from one time-step to another.  However, if we make
the ``time averaged approximation'' (TAA) whereby one assumes that at each
time-step the threshold distribution can be approximated by its
time-averaged value, then $p(s|\eta) = \delta(s(\eta)-s)$ where $s(\eta)$
is just
\begin{equation}
s(\eta) = \int_0^\eta p_{{\rm thresh}}(x) \>{\rm d}x.
\label{sofeta}
\end{equation}
The avalanche size distribution then becomes
\begin{eqnarray}
p_{{\rm aval}}(s) &=& \int_0^\infty \delta(s(\eta)-s)\, p_{{\rm
stress}}(\eta)\> {\rm d}\eta
= {p_{{\rm stress}}(\eta(s))\over p_{{\rm thresh}}(\eta(s))}\nonumber\\
&=& {1\over W\sigma} {\rm e}^{-\eta(s)/\sigma}
(f + {\rm e}^{-\eta(s)/\sigma})
\label{paval}
\end{eqnarray}
where we have used Eqs.~(\ref{pthresh}) and~(\ref{sofeta}).  We can
calculate the stress level $\eta(s)$ corresponding to an avalanche of size
$s$ from the same two equations, which give
\begin{eqnarray}
s &=& \biggl[\log{1+f{\rm e}^{\eta/\sigma}\over1+f}\biggr] \bigg/
\biggl[\log{1+f{\rm e}^{1/\sigma}\over1+f}\biggr]\nonumber\\
&\approx& \sigma\log (1+f{\rm e}^{\eta/\sigma})-\sigma f
\label{sapprox}
\end{eqnarray}
for ${\rm e}^{-1/\sigma} \ll f \ll 1$ and $\sigma \ll 1$.  We can now
distinguish a number of different regimes.  For small avalanches, such that
$s\ll\sigma$, the logarithm on the right-hand side can be expanded giving
$s + \sigma f \approx \sigma f {\rm e}^{\eta/\sigma}$.  Substituting
into Eq.~(\ref{paval})
\begin{equation}
p_{{\rm aval}}(s) \propto [s + \sigma f]^{-2}\qquad\mbox{for $s\ll\sigma$.}
\end{equation}
This gives a flat avalanche distribution for small $s$ up to about
$s=\sigma f$, and then a power-law distribution for larger $s$ with
exponent $\tau=2$.  The approximation breaks down when $s\approx\sigma$,
giving way to a regime in which ${\rm e}^{\eta/\sigma} \sim {\rm e}^s$, and
hence the avalanche distribution falls off exponentially with $s$.  The
various regimes can clearly be seen in the numerical results presented in
Figure~\ref{size}, and the predicted cross-over points between
them agree well with the theory.

When $f$ decreases below ${\rm e}^{-1/\sigma}$, the approximations in
Eq.~(\ref{sapprox}) break down and instead it becomes valid to write ${\rm
e}^{\eta/\sigma} \approx 1 + s{\rm e}^{1/\sigma}$.  In this regime the
theory predicts a breakdown in the scaling, a phenomenon which is also seen
in the simulations.  Thus the reloading process, whose scale is set by $f$,
must be small but necessarily non-zero if we are to see power-law behavior
in the avalanche distribution.  Notice however that at precisely $f=0$ the
theory predicts a return to $\tau=2$ scaling, which is not seen in the
simulations, implying that the TAA breaks down in this regime because the
distribution $p(s|\eta)$ becomes too broad to be well approximated by a
$\delta$-function.

The physical principle behind the appearance of a power-law distribution
here is the interdependence of the avalanche and threshold distributions;
the avalanche distribution is a function of the particular distribution of
thresholds at any time, but the threshold distribution is itself produced
by the action of the avalanches.

\section{Connection with other models}
There are clear similarities between our model and the sand pile model, in
which sites also possess a certain threshold stress that they will
withstand without adjusting.  Furthermore we have a source term, the
reloading or aging fraction $f$ of agents which at each time-step loose
memory of their previously assigned thresholds.  This source term is
similar in effect to the addition of the single grains of sand in the sand
pile models.  There are however some important differences between our
model and the SOC models.  First, the stresses in our model are coherent,
rather than localized as they are in the sandpile.  Second, the agents are,
at least in the simplest versions of the model, entirely non-interacting.
In SOC models, it is the interactions which give rise to avalanches.  In
our model on the other hand the avalanches of simultaneously moving agents
arise because all the agents feel the same externally imposed stresses.
There is no causal connection between the events which comprise an
avalanche; each agent moves independently of the others.

Unlike other model systems for large scale fluctuations, such as the
Burridge-Knopoff~(BK) model~\cite{BUK} and the recycled version of the
Democratic Fiber Bundle Model~(DFBM)~\cite{DFBM}, the model presented here
does not make a clear distinction between small, finite-sized events, and
large ones whose size scales like the size of the system.  In the BK model,
for instance, the spectrum of event sizes contains two separate parts, one
composed of small events which scales as $s^{-2}$, and another composed of
the big events, which occur quasiperiodically.  The BK and DFBM models are
not statistically stationary, by contrast with our model whose dynamics
rapidly reaches a statistically stationary state.  Models such as BK and
DFBM also show ``foreshock'' events in which large avalanches are preceded
by smaller ones.  Our dynamics does not have foreshocks but does display
aftershock events, a phenomenon which we discuss in greater detail in next
section.

\section{Discussion}
Next, we would like to examine the potential relationship of our model to
processes occurring in real physical systems.  First we consider
earthquakes.  To begin with, we ignore spatial correlations and consider
the variables $x_i$ to be thresholds for movement at various points along a
fault.  The coherent stress $\eta$ is provided by long-wavelength
background noise from some external source, such as other distant tremors,
or movements in the deeper regions of the earth, and the reloading $f$ is
due to slow plastic deformation from tectonic movements of the crust.  As
we have seen, these elements alone lead directly to a power-law
distribution of earthquake sizes very close to the observed
Gutenberg-Richter law, without the need to invoke interactions between
neighboring parts of the fault.  That is not to say that such interactions
do not exist, only that they are not necessary to produce the observed
power law.  (Kagan~\cite{Kagan} has presented evidence of a fractal pattern
in the spatial distribution of earthquake activity, which is an indication
that interactions are a feature of the dynamics.  This however need not
lead us to conclude that these local interactions are necessary for
producing the observed size distribution of events.)

\begin{figure}
\begin{center}
\psfig{figure=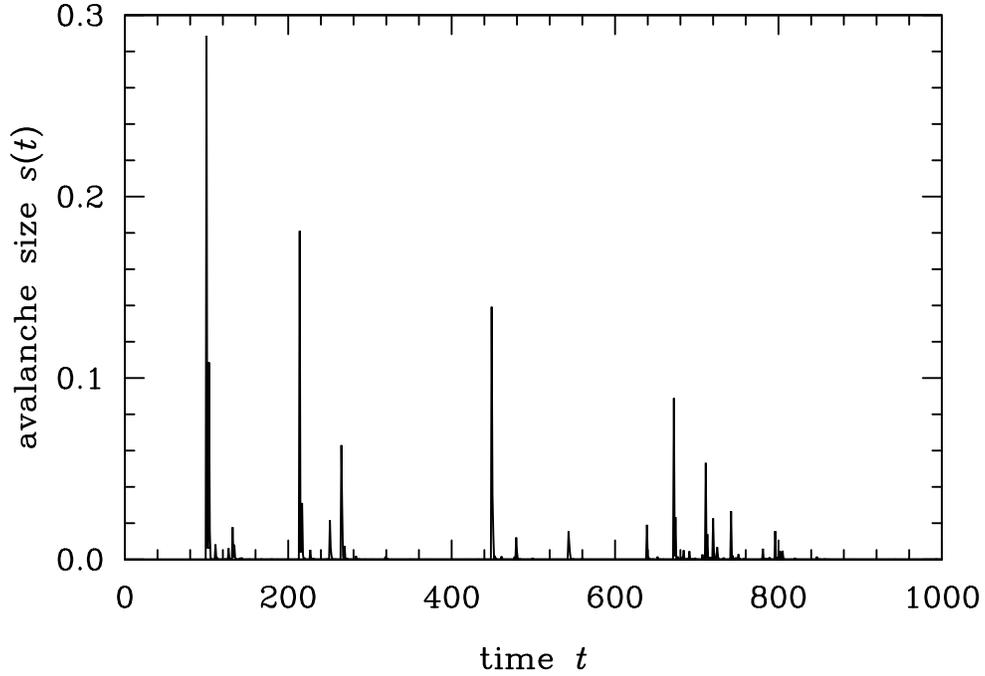,width=13cm}
\end{center}
\caption{Time series plot of avalanche sizes during a portion of a
simulation showing clear aftershocks.
\label{after}}
\end{figure}

\begin{figure}
\begin{center}
\psfig{figure=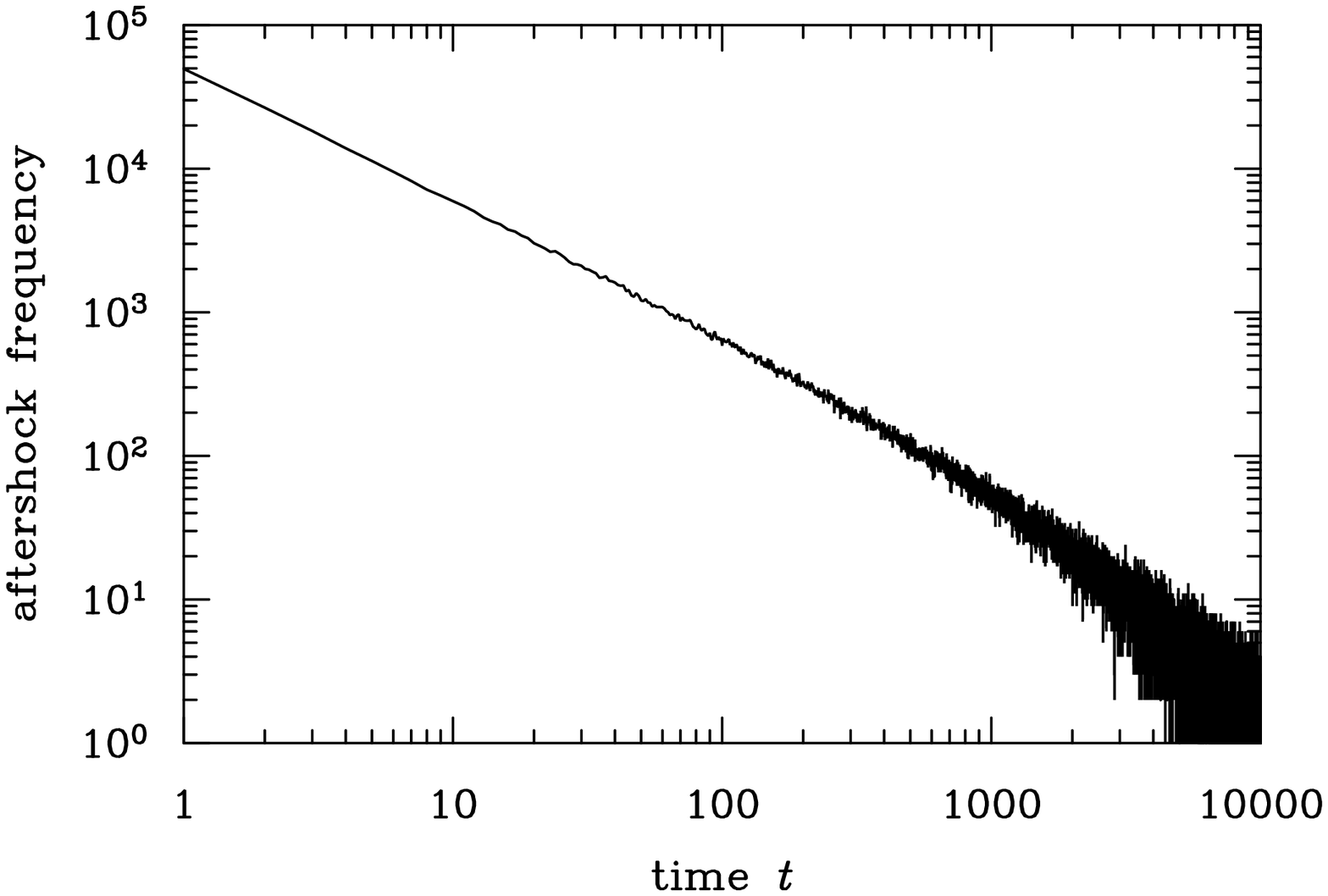,width=13cm}
\end{center}
\caption{Histogram of the time distribution of aftershocks following a
major avalanche.  The histogram follows a power law with an exponent close
to one (Omori's law).
\label{omori}}
\end{figure}

Another interesting feature of our model is that it shows clear
aftershocks.  The mechanism for these is straightforward.  When a large
avalanche takes place, a significant fraction of the thresholds in the
system are replaced with new, uniformly distributed ones.  Because of the
monotonic increasing form of the threshold distribution, this has the
effect of shifting the weight of the distribution downwards, increasing the
fraction of agents with low threshold for movement.  The result is that
subsequent stresses on the system have a larger-than-normal effect, and we
see an amplification of the usual level of ``background'' avalanches in the
aftermath of a particularly large event.  In Figure~\ref{after}, we show a
section of a time series of avalanches from one of our simulations, which
clearly displays this aftershock effect.  Notice that if we apply the
argument iteratively, we would also expect to see sequences of
``after-aftershocks'' following each of the aftershocks, a behavior which
is indeed evident in Figure~\ref{after}.  We have also measured the average
probability of getting an event of significant size in the aftermath of
another large one, and found that for small times its distribution goes
approximately as $t^{-1}$ (Figure~\ref{omori}).  A similar result is seen
in the data from real earthquakes, and is commonly referred to as Omori's
law~\cite{Omori}.

The $t^{-1}$ distribution can be understood as follows.  A large avalanche
will redistribute the thresholds of a large fraction of the agents in the
system uniformly across the interval of allowed values ($0\le x<1$ in this
case).  A subsequent stress of magnitude $\eta_1$ will remove all those
agents with $x<\eta_1$, and produce an aftershock extinction of a certain
magnitude.  In order to get another significant aftershock we now need a
stress $\eta_2>\eta_1$ in order to reach those agents which were not
affected by the first aftershock.  In general, if it took a time $t$ to get
the first stress, then it will on average, take as long again to get
another of the same magnitude, or an aggregate time of $2t$ until a stress
of size $\eta_2$ comes along.  Repeating the argument, it will take as long
again, or a total time of $4t$ to get the third aftershock, and so forth.
Given this exponential increase in the time intervals between these events,
it is not hard to show that the histogram of aftershock events should have
a $t^{-1}$ power-law form, regardless of the precise distribution of
stresses applied to the system.

Note that our mechanism is by no means the only way to obtain aftershocks.
An alternative mechanism has been proposed by Nakanishi~\cite{Hiizu} using
a Burridge-Knopoff-like model in which relaxation processes are introduced
by considering the geometry of stress redistribution following large
quakes.  As with the BK model, Nakanishi's model has a quasiperiodic
dynamics.

Second, let us compare our model with the results of recent studies of
one-dimensional rice piles by Frette~{\it{}et~al.}~\cite{Rice}.  In these
studies the experimenters found a frequency distribution of avalanche sizes
$s$ which was flat up to a certain fraction of the total size of the pile,
and then fell off as a power of $s$ for larger avalanches according to
Eq.~(\ref{powerlaw}), with a measured exponent of $\tau=2.1\pm0.1$.  A
similar behavior is seen in the simulation results from our model
(Figure~\ref{size}) which also display a flat distribution of avalanches up
to a certain fraction $\sim\sigma f$ of the total system size, and then a
power-law fall in avalanche frequency for larger sizes with exponent close
to two.  A possible interpretation of the experimental data then is that the
dynamics of the rice pile is one of avalanches produced by the interplay of
reloading with coherent stress.  The reloading $f$ could arise as a result
of newly added grains of rice, which tend to randomize the thresholds for
grains on the surface, and the stresses might come from the tumbling of new
grains as they are added to the pile.  The plateau in the avalanche
distribution for small sizes $s$ is then caused by rice grains which tumble
past a number of sites before coming to rest, but have only enough energy
to disturb the most unstable of those sites, and the larger events which
form the $\tau=2$ power law are the result of occasional larger stresses in
the tail of the distribution.  The one-dimensional nature of the system
ensures that all input disturbances propagate through a large portion of
the system, and thus may be treated as coherent.  In a two-dimensional
system this would not be the case, and the pile might well show entirely
different dynamics, either possessing a shallower power-law distribution
$\tau<2$, indicating perhaps that a true SOC dynamics is at work, or not
possessing a power-law distribution at all, indicating that coherent
driving forces are the only mechanism responsible for power laws in this
system.

Our model also makes quantitative predictions about the scaling of the
line between the two regimes in the avalanche distribution: the position of
the line should go like $N\sigma f$, the factor of the system size $N$
appearing when we shift from measuring avalanches as fractions of the
system size to measuring the total energy they release.  Scaling of
precisely this form with $N$ is indeed seen in the experiments.
Frette~{\it{}et~al.}\ also mention that simple scaling disappears when the
experiment is repeated with ``rounder'' rice.  We can explain this result
in terms of the narrower distribution of thresholds that round rice can
support, which corresponds to larger values of both $\sigma$ and $f$.

The results of these experiments have also been modeled by
Christensen~\cite{Frette} using a SOC model with interacting elements.
Clearly, there are aspects of the dynamics captured by their model which
are missing from ours, particularly geometrical effects concerned with the
spatial distributions of avalanches and the corresponding transport
properties of rice in the pile~\cite{Paczuski}.  However, because the
exponent $\tau$ is greater than 2, making $\langle s \rangle$ independent
of system size, we can expect these properties to be independent of the
largest avalanche events (though on the other hand, they should now depend
strongly on the position and nature of the crossover between the two
regimes of the avalanche distribution).  We suggest that the reverse is
also true, i.e.,~that the observed large avalanches could appear even in
the absence of long range spatial correlations.

One characteristic which does seem to distinguish our model from the SOC
alternatives is the existence of aftershock events.  It might therefore
might be profitable to investigate the existence of aftershock avalanches
in the experimental data, in order to make a quantitative distinction
between the two classes of dynamics.

Finally we would like to point out that models of the type introduced here
do not constrain $\tau$ to values close to $2$.  Although the values found
with the simple version of the model outlined in Section~II all lie
approximately in the range $1.8<\tau<2.4$, we have investigated other
variants on the model which produce values outside this range.  One
particularly interesting version is one in which we allow for the
possibility of there being many different kinds of stress on an agent.  We
suppose that agent $i$ is subject to $M$ independent types of stress, and
that it has a separate threshold for yielding to each one, making ${\bf
x}_i$ an $M$-dimensional vector quantity.  One then assumes that all $M$
components of ${\bf x}_i$ are to be replaced with new values every time any
one of the types of stress exceeds the corresponding threshold value.  In
the limit $M=1$, this model is clearly just the same as the version
discussed above, and for higher values of $M$ we continue to see a
power-law distribution of avalanche sizes, regardless of the nature of the
applied stresses.  However, the exponent of the power law becomes steeper
as the value of $M$ increases, and appears to approach 3
as $M$ becomes large.  (We have investigated the model numerically up to
$M=50$.)  It is interesting to note that stock market fluctuations show
power-law fluctuation distributions with exponents close to
$\tau=3$~\cite{Economic}.  One may speculate whether these so-called ``fat
tails'' in the distribution
are the natural response to the action of external stresses on the market
(of which there are indeed many).

\section{Conclusion}
To summarize, we have demonstrated that coherent noise in large systems
typically gives rise to intermittent behavior with an ``avalanche'' type
dynamics characterized by a power-law distribution of avalanche sizes with
exponent in the vicinity of $\tau=2$.  This value is similar to that
seen in a number of real systems, including rice piles and earthquakes,
suggesting that these systems may in fact be driven by external noise,
rather than self-organizing under the influence of short-range internal
interactions.  If one allows more elaborate types of stress on the system
one can obtain power laws with exponents as high as $\tau=3$.

We believe that the study of systems driven in this fashion by coherent
external noise may offer new interpretations of intermittent dynamics in a
variety of extended non-equilibrium systems in terms of a direct interplay
between small scale structures and long wavelength fluctuations in the
system.  Such systems might include not only the ricepiles and earthquakes
considered here, but possibly also extended chaotic systems such as
economics~\cite{Economic} and turbulence~\cite{Turbu}.

\begin{references}
 \bibitem{BTW}
 {\frenchspacing P. Bak, C. Tang and K. Wiesenfeld, Phys. Rev. 
Lett. {\bf 59} 381 (1987).}
 \bibitem{Quakes}
K. Christensen and Z. Olami, J. Geophys. Res. {\bf97}, 8729 (1992);
Phys. Rev. A {\bf46}, 1829 (1992).
 \bibitem{BS}
 {\frenchspacing P. Bak and K. Sneppen, Phys. Rev. Lett. {\bf71}, 4083
(1993); K. Sneppen, P. Bak, H. Flyvbjerg and M. H. Jensen,
Proc. Nat. Acad. Sci. {\bf92}, 5209 (1995).}
 \bibitem{NR}
 {\frenchspacing M. E. J. Newman and B. W. Roberts, Proc. Roy. Soc. B
{\bf260}, 31 (1995).}
 \bibitem{Fires}
 {\frenchspacing K. Chen, P. Bak, and M. Jensen, Phys. Lett. A {\bf149},
 207 (1990).}
 \bibitem{Review}
 {\frenchspacing M. Paczuski, S. Maslov and P. Bak, 
Phys. Rev. E {\bf 53}, 414 (1996).}
 \bibitem{deBoer}
 {\frenchspacing H. Flyvbjerg, K. Sneppen and P. Bak, Phys. Rev. Lett.
{\bf 71} 4083 (1993).\\
J. de Boer, B. Derrida, H. Flyvbjerg, A. D. Jackson and
T. Wettig, Phys. Rev. Lett. {\bf73}, 906 (1994).}
 \bibitem{Footnote}
In fact it could be argued that $\tau=2$ is the upper ``critical'' value
for 1D SOC models because the spatial organization on a length scale $L$ is
due to avalanches of size larger than $L$.  The probability for these to
appear is of order $1/L^{\tau-1}$, and, since they affect $L$ sites their
weighted contribution to ordering on length scale $L$ is $L/L^{\tau-1}$.
This average organization on length scale $L$ should be compared with the
disorganization caused by non-overlapping avalanches appearing with
frequency $\propto 1-1/L^{\tau-1}$.  Criticality demands that organization
should be larger than disorganization for $L\to\infty$, implying that
$\tau<2$.
 \bibitem{GRlaw}
 {\frenchspacing B. Gutenberg and C. F. Richter, Ann di Geofis. {\bf9}, 1
(1956); H. Kanamori and D. L. Anderson, Null. Seismol. Soc. Am.  {\bf65},
1073 (1975).}
 \bibitem{Rice}
 {\frenchspacing V. Frette, K. Christensen, A. Malte-S{\o}rensen, J. Feder,
T. J{\o}ssang and P. Meakin, Nature {\bf379}, 49 (1996).}
 \bibitem{BUK}
{\frenchspacing R. Burridge and L. Knopoff, Bull. Seismol. Soc. Am. {\bf 57},
341 (1967).\\
J.M. Carlson, J.S. Langer, B. Shaw and C. Tang, 
Phys. Rev. A {\bf 40} 6470 (1991).}
\bibitem{DFBM}
 {\frenchspacing D. Sornette, J. Phys. I, France {\bf 2}, 2089 (1992).}
 \bibitem{Kagan}
 {\frenchspacing Y. Y. Kagan, Physica {\bf D77}, 160 (1994).}
 \bibitem{Omori}
 {\frenchspacing F. Omori, J. Coll. Sci. Imp. Univ. Tokyo {\bf7}, 111
(1894); T. Utsu, Geophysical Magazine (Japan Meteorological Agency, Tokyo)
{\bf30}, 521 (1961).}
\bibitem{Hiizu}
{\frenchspacing H. Nakanishi, Phys. Rev. A {\bf 46} 4689 (1992).}
 \bibitem{Frette}
{\frenchspacing K. Christensen, A. Correl, V.Frette, J. Feder and
T. J{\o}ssang, preprint 1996 (cond-mat/9602067).}
\bibitem{Paczuski}
{\frenchspacing M. Paczuski and S. Boettcher, preprint 1996
(cond-mat/9603085).}
\bibitem{Economic}
{\frenchspacing P. Bouchaud and D. Sornette, Preprint Univ. Nice-Sophia,
Antipolis; R. Mantegna and H.E. Stanley, Nature {\bf 376}, 46 (1995)}.
\bibitem{Turbu}
{\frenchspacing B. Castain, Y. Gagne and E. Hopfinger, 
Physica D {\bf 46} 177 (1990); B. Tabeling, G. Zocchi, F. Belin, 
J. Maurer and H. Willaime, Phys. Rev. E {\bf 53} 1613 (1996).}
 \end{references}
\end{document}